# Tails & Tor

… and other tools for Safeguarding Online Activities

Authors:

Stephanie Abraham
Tyler Silva
Robert Decourcy
Jim Cardon

# Forensic Investigations

"**Tails and Tor** are well known tools used to legitimately protect privacy or to obfuscate criminal acts. **Explain how these tools work.**"

"Detail ways in which an investigator can **acquire Tor evidence** and include **limitations with this type of analysis**. Detail ways in which investigators can still **track users on these private networks.**"

"Identify **other tools** to hide the identity of criminal actors. Include the **latest trends used by hackers and criminals** to cover their activity and identify **new technologies, currently being used**."

# Table of Contents

# What is Tails?

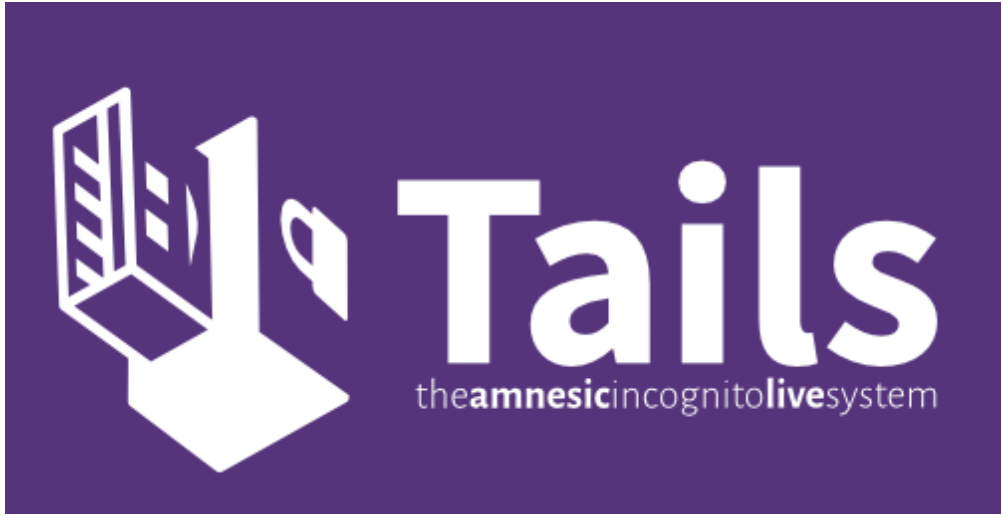

Tails ("The Amnesic Incognito Live System") is built as a self-contained operating system, intended to boot securely from a read-only flash drive or DVD. It is based on the Debian Linux distribution. It is similar in function to the earlier Live OS distribution of Incognito, which was built on a Gentoo Linux.

By running as a Live OS from removable, read-only boot media, Tails provides a level of privacy by ensuring that no traces are left on the workstation it is run on. When the user requires non-volatile storage, that storage would normally be encrypted using Linux Unified Key Setup (LUKS) disk encryption.

Since analysis of a local Tails workstation will not be expected to provide useful detail on the user activity, an investigator will need to rely mostly on network and cloud evidence.

Applications installed as part of the Tails distribution include the Tor Browser for anonymous web browsing based on Mozilla Firefox, Pidgin with OTR for encrypted instant messaging, and Claws with GnuPG for encrypted email and newsgroup access. "Dark Web" access is provided via both Onion Routing (Tor) and the Invisible Internet Project (I2P).

The OS is better known for its anonymous nature with valuable tools that is free and opensource. The Tails group has intricate reasons for why they chose this software. The problems branch from different exploits that are waiting to occur in different add-ons and JavaScript. On their website, the Tails working group tells of how their website would be flagged if they used their own web certifications, modern surface browsers such as Mozilla Firefox, Safari, and Internet Explorer would flag it, and not many people would trust their services, so in order to avoid said conundrum they rely on HTTPS / SSL certificates which is recognized by most, if not all browsers.

Tails offers intense anonymity tools such as: Pidgin (IM chat client), Icedove - and E-mail client, Electrum - a Bit Wallet service, and MAT (removes). These are all for both protection, and also can be used in a promiscuous way. The Tails group strongly believes in laissez-faire democracy based on all the tools they hand their customers, and the potential these tools posses.

## Tails Evidence

There are plenty of holes to the anonymity of Tails since it's an entire operating system. Tails is Debian based, so it is required to follow the major functions and protocols of any Debian machine, but remain more anonymous about it. Using a simple Linux command, I was able to view the different logs that the machine was recording. Taking a look at this data is crucial because an investigator can view the actions that a suspect took; assuming they used the Tails Operating System in conjunction with Tor.

```
amnesia@amnesia:~$ cd  var/log
bash: cd: var/log: No such file or directory
amnesia@amnesia:~$ ls
Desktop   Documents   Downloads   Music   Pictures   Public   Templates   Tor Browser   Videos
amnesia@amnesia:~$ cd  ..
amnesia@amnesia:/home$ cd  ..
amnesia@amnesia:/$ ls
bin    etc           initrd.img.old   media   proc   sbin   tmp   vmlinuz
boot   home          lib              mnt     root   srv    usr   vmlinuz.old
dev    initrd.img    live             opt     run    sys    var
amnesia@amnesia:/$ cd  var/log
amnesia@amnesia:/var/log$ ls
alternatives.log   btmp          faillog              hp              live              tor
apt                cups          fontconfig.log       htpdate.log     live-persist      wtmp
boot.log           dmesg         fsck                 i2p             macchanger.log    Xorg.0.log
bootstrap.log      dpkg.log      gdm3                 lastlog         speech-dispatcher
amnesia@amnesia:/var/log$
```

This screenshot shows the exploration of the different log files that Linux records. The only issue that an investigator may have with Tails is that in order to conduct investigations, the investigative must have the root password as well as administrative access to the computer.

## Difficulties

Conducting an investigation on a Linux machine is a difficult task, that can't have any guarantees without the root access to that device. I was denied access because I couldn't define my administrative privileges on my machine. While scrolling through these directories I began to see why people prefer to use Tails over many other distros and flavors of Linux.

```
amnesia@amnesia:/var/log$ cd  tor
bash: cd: tor: Permission denied
amnesia@amnesia:/var/log$ cd  i2p
bash: cd: i2p: Permission denied
amnesia@amnesia:/var/log$ cd gdm3
amnesia@amnesia:/var/log/gdm3$ ls
ls: cannot open directory .: Permission denied
amnesia@amnesia:/var/log/gdm3$ cd ..
amnesia@amnesia:/var/log$ cd htpdata.log
bash: cd: htpdata.log: No such file or directory
amnesia@amnesia:/var/log$ open htpdate.log
Couldn't get a file descriptor referring to the console
amnesia@amnesia:/var/log$ open fontconfig.log files
Couldn't get a file descriptor referring to the console
amnesia@amnesia:/var/log$ open fontconfig.log
Couldn't get a file descriptor referring to the console
amnesia@amnesia:/var/log$
```

Tails is incognito by design and look which another issue that allows people to roam the dark web on public Wi-Fi locations, and have investigators take a look at their IP Address while he's scrolling. Tails comes equipped with tools to spoof a Mac Address in my little "Red Team, Blue Team" experiment.

The only sure possible way to pin a suspect with the crimes they committed would have to be find the True MAC Address of the machine they are currently using. In summary the answer to the question "is it possible to catch a suspect on their device running Tails Linux?" yes. "Is it difficult?" extremely.

# What is Tor?

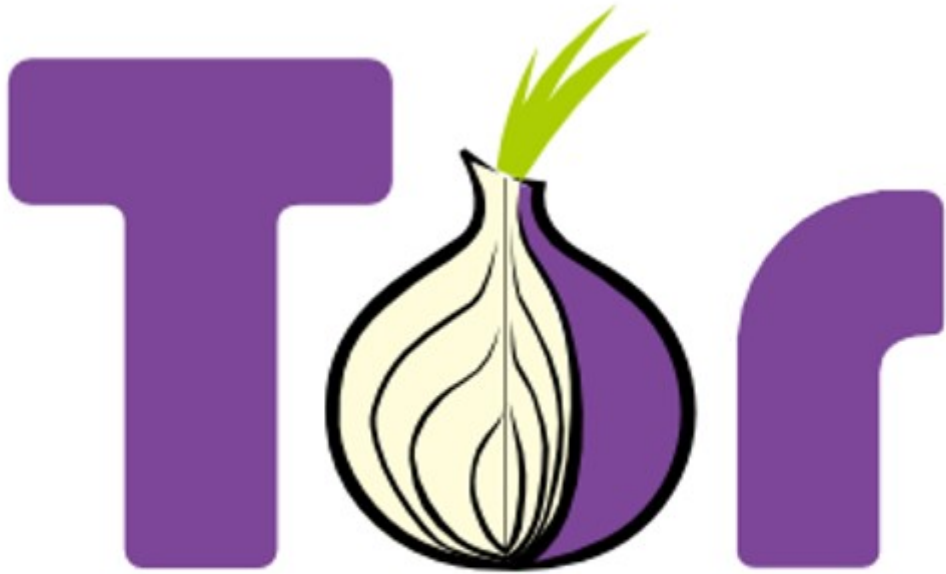

Tor is essentially an overlay that runs on top of the public Internet, allowing users greater end-to-end anonymity via an Onion Routing design. It also provides a method of hosting "Dark Web" services that are not reachable via the standards based Domain Naming System. Onion Routing was first developed by the US Naval Research Lab, and is now managed by the non-profit Tor Project. The Tor Project is still financed in the most part by the US Government.

Tor is named after the original project name "The Onion Router" although it is no longer an acronym and is simply called Tor. It has been aptly named for the way that data is encrypted and relayed through "layers". Data packets are encrypted, and the packet header is stripped. This is a very important step since the header contains information about the sender. Tor then encrypts the remaining information in a wrapper. This process is typically repeated three times with each layer containing information about the next relay destination. Each relay that receives the packet only has the key to decrypt the next wrapper and destination IP of the next relay. This continues until the packet arrives at its final destination.

The dark web itself is legitimately useful to many law abiding citizens, however it is a very conducive environment for illegal activities and services. Many of these range from purchasing illegal weapons and drugs, child pornography, and hitmen. There are also many useful tools for anyone that is trying to stay anonymous. These include untraceable email systems, whistleblower sites, forums, libraries of books banned in certain countries, and file storage.

## Tor Evidence

**How can an investigator acquire Tor evidence? What are some limitations of Tor analysis? How can investigators still track users on these private networks?**

Despite the number of illegal activities that occur in the dark web, it has taken the FBI an enormous amount of resources to crack down on them. A typical VPN is generally regarded as secure for most uses, however website fingerprinting revealed a 90% accuracy when attempting to identify HTTP packets, while Tor had a 2.96% accuracy.

There are a few methods which can be used to compromise the anonymity of the Tor network, although many of them do not completely break it alone. The Heartbleed bug only resulted in minimal downtime for less than a quarter of the relays and no breaches in security. This was partly because Tor uses two sets of keys.

Tor is a distributed trust design meaning that compromising a single relay or node will not affect the overall security of the network. Controlling a large portion of the network would still present problems.

Despite this, some of Tor's security features can be exploited. For example, since clients and hosts have zero authentication between them, man-in-the-middle attacks are possible for torrent tracker information. It has been confirmed that the FBI hired Carnegie Mellon to break Tor's anonymity in pursuit of the Silk Road 2.0 (drug marketplace), spending over $50,000 in

computer time alone. Carnegie Mellon has declined any question about the methods that were used.

# Tools, Trends, & Technology

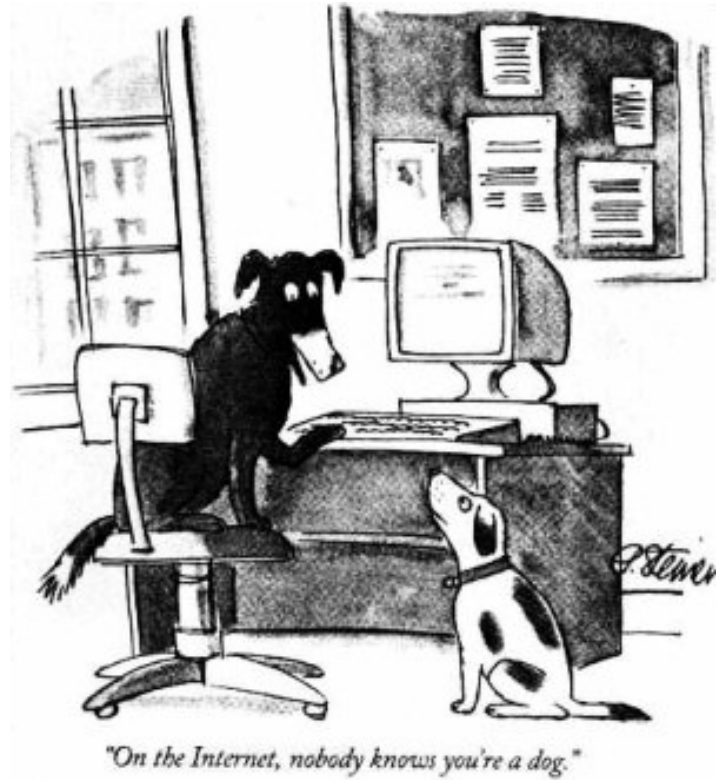

*The New Yorker*; July 5, 1993 [1]

One of the great strengths of the Internet is that users are able to protect their identity, and their actions, from being tracked by other parties. But from a law enforcement perspective, these capabilities may allow criminals to commit illegal activities in ways which are difficult for a forensics investigator to unravel.

## Fake and Anonymous Identities

Criminals and criminal enterprises may create fake online personas in order to help carry out their activities. Last year, Dell SecureWorks Counter Threat Unit produced a report on networked LinkedIn accounts which were used as part of a social engineering attack on individuals in the mobile telecom industry [2]. The goal of this type of bogus identity is to collect information on, and build a profile of, the targeted users. Once a profile of the target is built, it is easier to perform spear-phishing, whaling, and other pinpoint social engineering attacks.

Creation of a fake identity is rather straightforward, using simple online tools. To demonstrate, we generated details for a fake identity "Kenneth Lane", assisted by FakeNameGenerator.com. From there, we leveraged a disposable email address on ThrowAwayMail.com to create a soft telephone number with TextMe iOS app, and to also create a legitimate looking gmail account. Along with an online SMS receiver, we created Facebook and LinkedIn accounts for Ken as a Northeastern University graduate working at Berkshire Global Partners with past history at UBS. After sending a batch of 75 connection requests to LinkedIn users in those educational and employment networks, we began growing our connections. Overnight, we had 50 connections (a 66% hit rate!), many from UBS, with titles ranging from VP Financial Service, to Board of Advisors, to Senior Manager. Beyond some expected recruiter connections, requests, and direct messaging, Ken also received a more surprising "Hi Kenneth, Congrats on the new gig! Hope you're doing well." message from a "fellow" alum. This type of open acceptance of connections from complete strangers is what forms the basis for successful social engineering attacks.

Not all anonymous activity requires a fully developed fake persona. Apps like Kik and HitMeUp! are known to be used by online predators for creating anonymous messaging accounts [3]. By not requiring verifiable phone number authentication, accounts can be created that are not easily traceable to an actual user. Browsing on HMU! for others with shared interests, or by exchanging Kik information on anonymous location based services like Yik Yak, there is a documented ease for pedophiles to exchange pictures or create child exploitation situations [4].

| Name<br>Address<br>Birthdate | fakenamegenerator.com | Kenneth R. Lane<br>193 Adams Ave, Newton, MA 02465<br>September 26, 1983 |
|---|---|---|
| Disposable eMail | ThrowAwayMail.com<br>GuerrillaMail.com | huphulidre@thraml.com<br>klane617@sharklasers.com<br>3o1rkj+8lu65lt3q56ss@sharklasers.com |
| Burner Phone Number | sellaite.com/smsreceiver<br>TextMe iOS App | +372-5820-2935 (Estonia)<br>+1-617-213-9898 |
| "Legitimate" eMail | gmail.com | kenlane617@gmail.com |

| LinkedIn Account | linkedin.com | kenneth-lane-82137a11a<br>Northeastern University, Class of 2005<br>Berkshire Partners & UBS |
|---|---|---|
| Facebook | facebook.com | Ken Lane, id=100011848471323 |
| Anonymous Messaging | HitMeUp!<br>Kik | klane617<br>kenlane617 |

## Secure Communication

Besides masking of identities, tools are available for criminals to securely communicate. While apps like Kik do provide some basic in-flight encryption, the keys are still stored on central servers. After being hit with a public Man-in-the-Middle attack, the WhatsApp moved to an end-to-end encryption model using device-distributed keys[5]. But after WhatsApp was acquired by Facebook, many users moved to Threema and other alternate secure messaging services[6]. Better options might be the use of open source messaging clients like Pidgin with an encryption plugin like Off The Record. This is a way to ensure that the key based encryption is fully under the control of the user [7]. A user-controlled, certificate based encryption model offers assurances for the authenticity, integrity, and confidentiality of the messaging. Similarly, email or file transfer information can be protected using open source GnuPG or OpenPGP tools, which also give the users ownership of their public/private keys pairs. Edward Snowden's exposure of government surveillance programs also exposed how OTR and PGP, as well as some online messaging providers like Zoho, provide great difficulty for intelligence services to decrypt [8]. But key size is critical to keep the data confidential. Last year a paper out of University of Pennsylvania showed how Amazon Web Services could be used to cheaply and quickly brute force 512-bit RSA encryption [9]. And it is very likely that the NSA has pre-computed the prime keys needed to decrypt at least 1024-bit encryption in real time [10].

Of course, there is still risk of unannounced 0-day vulnerabilities being used to decrypt this traffic, but this risk exists in any digital tools and is actually mitigated somewhat by the nature of the open source community. A full-disclosure policy is not something which many vendors or large corporations like, but it does provide smaller players with knowledge of new vulnerabilities as soon as they are found in the field [11].

Looking beyond the encryption of messaging in flight, full-disk encryption is also becoming mainstream for protection of data at rest, this poses a significant hurdle for criminal and forensic investigators. BitLocker, FileVault, and LUKS provide full disk encryption on the desktop, while Android Marshmallow and iOS provide full encryption for mobile devices. The San Bernardino terrorism case shows how difficult it is for law enforcement to break this type of encryption. While the iPhone 5c used in by the terrorists in this case has apparently been cracked, the 5s and later hardware are still not known to be exploitable. It is likely that this case will result in additional federal legislation to strengthen the existing CALEA requirements [12].

## Obscurity and Anonymous Activities

As discussed earlier in this paper, use of a Live OS like Tails, with a Tor browser across an onion routing network - or the Invisible Internet Project (I2P) - provides a high level of anonymity for a user. Other ways of masking network location would be by using online proxy services such as vip72, megaproxy, and anonymizer. During the 2008 presidential elections, David Kernel used the CTunnel proxy during a hack of Sarah Palin's email account [13]. Among several mistakes Kernell made, the proxy (which he exposed in screen shots) logged his own IP address. Had he been using Tor, this exposure would have only led investigators to the exit router in the Tor network, not directly to him. Kernell also left a traceable email address on his 4chan posting, rather than a disposable address which would have been more difficult to track.

An alternative way for sensitive information to be released is through the tool known as SecureDrop, funded by the Freedom of the Press Foundation. This provides an anonymous way for whistleblowers to submit information to the media. Edward Snowden is a prime example of how Tor based tools like SecureDrop could provide greater anonymity, but Snowden also shows how not all disclosures to the media are considered "whistle blowing"; instead his actions are classified by many as treason or at least criminal. Also developed by Micah Lee and the Freedom of the Press Foundation is a similar tool known as OnionShare, which allows for more generic anonymous sharing of files across temporary, hidden, Tor services[14].

Obscure communication channels are also available on the internet. For example, use of gaming consoles could provide a method of communicating between bad actors which is not traditionally monitored or investigated. There was discussion this past fall that PS4 Play Station Network may have played a role in communication between the Paris terrorists, but this has not been publicly confirmed. Just prior to the attack, comments from Belgium's interior ministry mentioned the PS4 as an example of a communication channel which lacks surveillance ability, but this network was never directly tied to the attacks which occurred after his warning [15].

## More Traditional Tools

Criminal activities also occur with more publicly known tools and attacks. Botnets are installed on unsuspecting user systems, generally through the use of broadly sent phishing emails with trojan viruses. A botnet army can be used for varying criminal purposes including DDoS attacks, spam mailing, data disclosures, and simple CPU vampiring[16].

Torrent and other peer-to-peer tools are often used for the widespread distribution of copyright protected files and stolen or confidential information. Applications like BitTorrent are not inherently anonymous, since the user's seeding IP address is known by the connected systems. Use of TorGuard or other proxies help obscure this information further[17].

Finally, secure and anonymous transfer of money is an important aspect of criminal activity. While traditional 419 advance fee scams - a $12.7B business in 2014 - primarily use Western Union money transfers [18], more modern criminals use decentralized, untraceable digital currency such as BitCoin. Recent cyber-ransom attacks in the healthcare industry began as traditional trojan horse attacks, to allow installation of ransomware encryption tools such as CryptoLocker. Once business-critical files were encrypted, BitCoin payment was demanded in return for the keys needed to unlock the systems [19]. BitCoin and other cryptocurrencies are used extensively on the dark web, as demonstrated by sites like Silk Road (currently "Reloaded" after several iterations and years of arrests). Of course, sometimes the criminals themselves are targeted, with millions of dollars' worth of BitCoins being stolen from Silk Road 2.0 escrow accounts in 2014 through a transaction malleability attack [20].

# Dark Pools

## What Are Dark Pools?

**An interview with Suzette Civaro, Managing Director at the New York Stock Exchange (NYSE).**

### What are Dark Pools, and how do they function?

"Dark Pools are anonymous, private exchanges used for trading, they function the same as regular stock exchanges but are not visible to the public. Dark Pool trading activity occurs directly between two parties without the use of an exchange."

The NYSE's modus operandi is "we guarantee a safe and transparent trading venue for buyers and sellers to find each other and trade with each other publicly on our trading platform."

Dark Pools represent secret and anonymous trading which is the complete opposite of what the NYSE stands for."

### How does that affect the NYSE and other businesses?

Two companies can mutually decide to anonymously do a trade off the exchange in a "dark pool" for large transactions i.e., $1,000,000 dollars +. This will curtail any suspicious trading activity that might negatively impact the stock from making future sales more difficult."

### Can you think of any practical ways an investigator can help?

"Investigators help uncover any dark pool operators who secretly offer high speed traders special order types that gave them an unfair advantage over other subscribers."

Dark pools provide secrecy, which can pose problem and some investors have faced allegations from the Securities and Exchange Commission, that they've treated customers unfairly. NYSE promotes "trade-at rule" that would keep more U.S. stock dealing on public exchanges run. Large investors want anonymous platforms to trade on,

Much like the dark web, Dark Pools allow a similar form of anonymity. Large Investors use dark pools to  buy and sell large shares without tipping off traders who otherwise will drive prices up or down against them. Controversy: A high-frequency trader with access to a dark pool will know when the official best price differs from the actual market price. Plans by the New York Stock Exchange to limit trading in so-called "dark pools" during an industry-wide experiment aimed at boosting trading in small-cap stocks have been put under review by the U.S. Securities and Exchange Commission. NYSE's proposed rules would limit the ability for trading firms to use one of the exceptions, the SEC said in its filing.

# References

### Tails & Tor

### Tools, Trends, & Technology